\documentclass[12pt] {article}

\usepackage{amssymb,amsmath}
\usepackage[round]{natbib} % gives options like citet, citep 
\usepackage{color}
\usepackage{graphicx}%[dvips]

\newcommand{\bal}[1]{\begin{align} \label{#1} }
\newcommand{\beq}[1]{  \begin{equation} \label{#1} }  
\newcommand{\eeq}{     \end{equation}}  

\renewcommand{\appendix}{
  \setcounter{section}{0}\renewcommand{\thesection}{\Alph{section}}
  \section*{Appendix} 
}

\def\Appendix#1{
  \setcounter{equation}{0}
  \renewcommand{\theequation}{\thesection.\arabic{equation}}
  \section{#1}
}

\allowdisplaybreaks%[1]
\newtheorem{thm}{Theorem}
\newtheorem{lem}{Lemma}

\newcommand{\rf}[1]{(\ref{#1})}

\def\bd#1{\mbox{\boldmath$\displaystyle\mathbf{#1}$} }
\def\tens#1{\mathbb{\,#1}}	   %tensor in 3D
 
\def\diag{\operatorname{diag}} 
\def\sym{\operatorname{sym}} 

\def\singlespacing{\baselineskip=13pt}	

\usepackage[pdftitle={Quadratic invariants of elastic moduli },
              pdfauthor={Andrew N. Norris},
              pdfkeywords={elasticity,invariants,isotropic,anisotropic}, 
              letterpaper=true,
              pdfpagetransition=Dissolve]{hyperref} 

\def\rev#1{#1}	   % for changes 

\begin{document} %%%%%%%%%%%%%%%%%%%%%%%%%%%%%%%%%%%%%%%%%%%%%%%%%%%%%%%%%%%%%%%%%%%%%%%%%

\pagestyle{myheadings}\markright{{\sc  Norris }  ~~~~~~Quadratic invariants }%\today}
\singlespacing%\doublespacing

\title{
\textcolor{blue}{Quadratic invariants of   elastic moduli }% under SO(2)}  
}

\author{Andrew N. Norris\\ \\    Mechanical and Aerospace Engineering, 
	Rutgers University, \\ Piscataway NJ 08854-8058, USA \,\, norris@rutgers.edu 	
}	\date{}

\maketitle

\begin{abstract}

A quadratic invariant is  defined as a quadratic form in the elements of a tensor that remains invariant under a group of coordinate  transformations.  It is proved that there are 7 quadratic invariants of the 21-element elastic modulus tensor under SO(3) and 35 under SO(2).   This answers  \rev{some} open questions raised by   \citet{Ting87} and \citet{Ahmad02} in this journal. 

\end{abstract}

\section{Introduction}

The tensor of elastic moduli $c_{ijkl}$ is known to possess two linear invariants under arbitrary proper orthogonal coordinate transformations, or SO(3):
\bal{066}
A_1& = c_{ijij} =c_{11}+c_{22}+c_{33}+ 2( c_{44}+c_{55}+c_{66} ), 
\nonumber \\
A_2& = c_{iijj} =c_{11}+c_{22}+c_{33}+ 2( c_{12}+c_{23}+c_{13}).
\end{align}
\cite{Ahmad02} presented four quadratic invariants under SO(3), the first two of which were reported by \cite{Ting87},  
\beq{067}
B_1 = c_{ijkl}c_{ijkl}, \qquad 
B_2 = c_{iikl}c_{jjkl}, \qquad 
B_3 = c_{iikl}c_{jkjl}, \qquad 
B_4 = c_{kiil}c_{kjjl}.  
\eeq
Ahmad demonstrated that the seven quadratic invariants $\{A_1^2,\, A_2^2,\, A_1A_2,\,
B_1,\, B_2,\, B_3,\, B_4\}$ are independent but did not show completeness.  It is clear that the following is an eighth invariant, 
\beq{080}
B_5 = c_{ijkl}c_{ikjl},
\eeq
although it is not so obvious whether or not it is independent of the other seven.  
We will prove  that there are at most seven independent quadratic invariants under SO(3), and  that the 
 seven identified by Ahmad  form a complete basis.  In particular,  $B_5$ is a linear combination of this basis, specifically   (see the Appendix)
 \beq{081}
 B_5 =  \frac12 A_1^2 + \frac12 A_2^2 - A_1A_2  + B_1 - 2B_2 + 4B_3 - 2 B_4  . 
\eeq
Consequently, every  fourth order elasticity tensor satisfies this identity:
\beq{-93}
2c_{ijkl}  (c_{ijkl}  - c_{ikjl})
+ ( c_{iijj}-c_{ijij})^2 - 4( c_{ijkk}-c_{ikjk})( c_{ijll}-c_{iljl})=0.
\eeq

It is well known that there are 5 linear invariants under rotation about an axis,  or SO(2).  Taking the axis as ${\bd e}_3$ these are  
\beq{9103}
L_1=c_{11}+c_{22}+2 c_{66},\quad L_2=c_{44}+c_{55},  \quad
L_3=c_{11}+c_{22}+2 c_{12},\quad L_4=c_{13}+c_{23},\quad L_5=c_{33} .
\eeq
\cite{Ahmad02} listed 17  quadratic invariants for SO(2):
\bal{9140}
E_{1}&=c_{34}^2+c_{35}^2  , \nonumber \\ 
E_{2}&=(c_{15}+c_{25})^2+(c_{14}+c_{24})^2   , \nonumber   \\
E_{3}&=(c_{15}+c_{46})^2+(c_{24}+c_{56})^2  , \nonumber \\ 
E_{4}&= (c_{14}+c_{24})c_{34} + (c_{15}+c_{25})c_{35}    , \nonumber   \\
E_{5}&= (c_{15}+c_{46})c_{35} + (c_{24}+c_{56})c_{34}   , \nonumber \\
E_{6}&=(c_{15}+c_{25}) (c_{15}+c_{46}) +(c_{14}+c_{24}) (c_{24}+c_{56})  , \nonumber \\ 
E_{7}&=c_{13}^2+c_{23}^2 +2 c_{36}^2  , \nonumber \\ 
E_{8}&=c_{44}^2+c_{55}^2 +2 c_{45}^2  , \nonumber \\
E_{9}&=(c_{11}+c_{12})^2+(c_{12}+c_{22})^2  +2 (c_{16}+c_{26})^2  , \nonumber \\
E_{10}&=c_{13} c_{55}+c_{23} c_{44}+2 c_{36} c_{45}   \\
E_{11}&= (c_{11}+c_{12})c_{13}  + (c_{12}+c_{22})c_{23} +2  (c_{16}+c_{26})c_{36}  , \nonumber \\
E_{12}&= (c_{11}+c_{12})c_{55} + (c_{12}+c_{22})c_{44} +2(c_{16}+c_{26}) c_{45}   , \nonumber \\
E_{13}&=c_{11}^2+c_{22}^2 +2 c_{12}^2 +4 (c_{16}^2+c_{26}^2 +c_{66}^2)  , \nonumber \\
E_{14}&=c_{14}^2+c_{24}^2 +c_{15}^2+c_{25}^2 +2 (c_{56}^2+c_{46}^2 )  , \nonumber \\
E_{15}&= (c_{44}-c_{55})c_{36} + (c_{13}-c_{23}) c_{45}  , \nonumber \\
E_{16}&=(c_{13}-c_{23}) (c_{16}+c_{26}) - (c_{11}-c_{22})c_{36}  , \nonumber \\  
E_{17}&=(c_{44}-c_{55}) (c_{16}+c_{26}) + (c_{11}-c_{22})c_{45}.  \nonumber
\end{align}
\cite{Ting87} had previously reported 15 invariants, and these can be shown \citep{Ahmad02} to be contained in Ahmad's  larger set.  
Ahmad also demonstrated that the 32 quadratic invariants formed from the 15 quadratic combinations of $L_1, \ldots, L_5$ plus the 17  invariants $E_1, \ldots, E_{17}$ are independent of one another.  The question of completeness remained open.  

The purpose of this paper is to finish the study initiated by \citet{Ting87} and subsequently expanded by \citet{Ahmad02}.   Two principal results are derived: first that the seven quadratic invariants under SO(3) identified by  \citet{Ahmad02} are indeed complete, and hence any quadratic isotropic invariant must be a linear combination of these.  Secondly, we prove 
that there are 35 quadratic invariants under SO(2).  A complete basis for the 35-dimensional space of quadratic invariants under SO(2) is formed  by the 32 of \citet{Ahmad02} augmented by the these three, 
\bal{303}
E_{18}&=  (c_{15}+c_{46}) c_{14} - (c_{24}+c_{56}) c_{25}  - c_{15}c_{56} +c_{24}c_{46},
\nonumber \\ 
E_{19}&= (c_{15}+c_{46}) c_{34} - (c_{24}+c_{56}) c_{35},
\\
E_{20}&= (c_{15}+c_{25}) c_{34} - (c_{14}+c_{24}) c_{35}. 
\nonumber
\end{align}

We consider quadratic forms on $\tens{E}$la,  
 the space of tensors of elastic moduli\footnote{To be precise, $\tens{E}$la comprises tensors that are positive definite, but that distinction is not necessary here as we are concerned with quadratic forms \emph{on} the moduli not defined \emph{by} the moduli.} 
which possess the underlying symmetries 
\beq{-2}
C_{ijkl}= C_{jikl},\qquad
C_{ijkl}= C_{klij},
\eeq
implying 21 independent elements, at most. The  general quadratic form on $\tens{E}$la is 
\beq{-1}
\phi = F_{ijklpqrs}C_{ijkl}C_{pqrs},
\eeq
where $F_{ijklpqrs}$ are the elements of an eighth order tensor.  Based on the properties \rf{-2} of the moduli, the elements of $F$ satisfy
\beq{-3}
F_{ijklpqrs} = F_{jiklpqrs},
\quad
F_{ijklpqrs} = F_{klijpqrs},
\quad
F_{ijklpqrs} = F_{pqrsijkl}. 
\eeq
Identifying quadratic invariants is therefore equivalent to describing the properties of the 
 eighth order tensor $F$, in particular the number of independent elements that survive under the group of transformations considered.   The problem is linked to that of finding the integrity basis for fourth order tensors. As noted in a review on tensor functions: ``Relatively little is known about representations of functions of tensors of order higher than two, for any of the transformation groups of interest in continuum mechanics"  
 \citep[p. 83]{Rychlewski1991}.   Considerable work has been done on this topic  for second order tensors since this is critical to the form of elastic strain energy functions \citep{Spencer71} (see \citet{Xiao96} for  recent developments on n=2).  However, this literature is not directly applicable since we are  concerned with tensors of higher order than normally considered.

 The  results derived here concern properties of 8$^{th}$ order symmetric tensors, and the number of constants they possess under transverse isotropy and isotropy, subject to the indicial symmetries \rf{-3}.

We begin in Section \ref{sec2} with a summary of the main results.  Although the focus of the paper deals with transformations caused by rotation, it will prove  useful to first consider invariance under reflection. Section \ref{sec3} introduces the notion of invariance under reflection about a plane, and quadratic invariants of the elastic moduli are derived in Section \ref{sec4} for transformations under one and two reflections.  Transformations under rotation about an axis  is then considered in Section \ref{sec5} where the  results are proved.  Finally,  issues of consistency and completeness are discussed in   Section \ref{sec6}.     

A note on notation:  The vector triad  ${\bd e}_i, i=1,2,3$ is an orthonormal basis in 3-dimensions.  The summation convention on repeated suffices is assumed.

\section{Summary of the principal results}\label{sec2}

The   results of the paper are  summarized in the form of two theorems, with the proofs given in the subsequent Sections. 

Define the 21-vector of elastic moduli
\beq{00}
c = \big( 
c_{11} \,  
c_{22} \,  
c_{33} \,    
c_{23} \,  
c_{13} \,  
c_{12} \,  
 c_{44} \,  
 c_{55} \,  
 c_{66} \,  
 c_{14} \,  
 c_{25} \,  
 c_{36} \,  
  c_{34} \,  
  c_{15} \,  
  c_{26} \,  
  c_{24} \,  
  c_{35} \,  
  c_{16} \,  
  c_{56} \,  
  c_{46} \,  
  c_{45}
  \big)^t, 
\eeq
then we have the following: 

\begin{thm}\label{thm1}
All quadratic invariants of $c$ under SO(3) are of the form $c^tFc$ where the 21$\times$21 symmetric matrix $F$ is of the form
\beq{3414}
F = \begin{pmatrix}
A_{9\times 9} & 0_{9\times 12}  \\
0_{12\times 9} & B_{12\times 12} 
\end{pmatrix},
\eeq
with symmetric matrices $A $ and $B$  defined by \rev{12 and 7 independent  elements, respectively}. These matrices have the form
\bal{501}
A &= \left(
\begin{tabular} {cccc cccc cccc}
       $f_{11}$ & $f_{12}$& $f_{12}$ & $f_{14}$ & $f_{15}$& $f_{15}$ &$f_{17}$     & $f_{18}$& $f_{18}$  
\\
&      $f_{11}$ & $f_{12}$& $f_{15}$ & $f_{14}$ & $f_{15}$& $f_{18}$ &$f_{17}$     & $f_{18}$ 
\\
& &    $f_{11}$ & $f_{15}$& $f_{15}$ & $f_{14}$ & $f_{18}$& $f_{18}$ &$f_{17}$      
\\
&&&    $f_{44}$ & $f_{45}$& $f_{45}$ & $f_{47}$ & $f_{48}$& $f_{48}$  
\\
&&&&   $f_{44}$ & $f_{45}$& $f_{48}$ & $f_{47}$ & $f_{48}$ 
\\
&&&&&  $f_{44}$ & $f_{48}$& $f_{48}$ & $f_{47}$  
\\
&S&Y&M&&&   $f_{77}$ & $f_{78}$& $f_{78}$  
\\
&&&&&&&  $f_{77}$ & $f_{78}$ 
\\
&&&&&&&& $f_{77}$  
\end{tabular} \right),
\\ & \nonumber \\
B &= \left(
\begin{tabular} {cccc cccc cccc}
       $b_{11}$ & 0& 0 & $b_{14}$ & 0& 0 &$b_{14}$     & 0& 0 & $b_{1\, 10}$     & 0 & 0 
\\
&      $b_{11}$ & 0& 0 & $b_{14}$ & 0& 0 &$b_{14}$     & 0& 0 & $b_{1\, 10}$ & 0 
\\
& &    $b_{11}$ & 0& 0 & $b_{14}$ & 0& 0 &$b_{14}$     & 0& 0 & $b_{1\, 10}$  
\\
&&&    $b_{44}$ & 0& 0 & $b_{47}$ & 0& 0 &$b_{4\, 10}$ & 0 & 0 
\\
&&&&   $b_{44}$ & 0& 0 & $b_{47}$ & 0& 0 &$b_{4\, 10}$ & 0 
\\
&&&&&  $b_{44}$ & 0& 0 & $b_{47}$ & 0& 0 &$b_{4\, 10}$  		%   6
\\
&&&&&&   $b_{44}$ & 0& 0 & $b_{4\, 10}$ & 0& 0 
\\
&&&&&&&  $b_{44}$ & 0& 0 & $b_{4\, 10}$ & 0 
\\
&&&&&&&& $b_{44}$ & 0& 0 & $b_{4\, 10}$   						%   9
\\
&S&Y&M&&&&&& $b_{10\, 10}$ & 0& 0  
\\
&&&&&&&&&&$b_{10\, 10}$ & 0   
\\
&&&&&&&&&&& $b_{10\, 10}$  
\end{tabular} \right),  \label{5011}
\end{align}
and   the 19 distinct elements  
\beq{01010}
f_{11},\,   f_{12},\,   f_{14},\,   f_{15},\,   f_{17},\,   f_{18},\,   f_{44},\,   f_{45},\,   f_{47},\,  f_{48},\,  
   f_{77},\,  f_{78},\,    b_{11},\,   b_{14},\,   b_{1\, , 10},\,     b_{44},\,  b_{47},\,  b_{4\, ,10},\,  b_{10\, ,10},
   \nonumber
\eeq
are arbitrary as long as they satisfy the 12 linearly independent conditions
\begin{subequations}\label{505}
\bal{505a}
2f_{1\, 1}+2f_{1\, 2} -4f_{1\, 5} + f_{4\, 4} -f_{7\, 7}  &= 0,
 \\
2f_{1\, 1}+2f_{1\, 2} -2f_{1\, 8} - f_{4\, 4}  -f_{4\, 7} &= 0, 
 \\
2f_{1\, 2} -f_{1\, 4} -f_{1\, 7}  &= 0, 
 \\
f_{1\, 4}+f_{1\, 5} -f_{4\, 5}  -f_{4\, 8} &= 0,
 \\
f_{1\, 7}+f_{1\, 8} -f_{4\, 8} -f_{7\, 8}  &= 0,
 \\ 
2f_{4\, 4} -2f_{4\, 5}  -b_{1\, 1} &= 0, 
 \\
2f_{1\, 4} -2f_{1\, 5} +b_{1\, 4} & = 0,
 \\
2f_{4\, 7} -2 f_{4\, 8} -b_{1\, 10} &= 0, 
 \\
4f_{1\, 1} -2f_{1\, 5}  -2f_{1\, 8} - b_{4\,4} & = 0, 
 \\
4f_{1\, 2} -2f_{1\, 5} -2f_{1\, 8}  + b_{4\,7} &= 0, 
 \\
2f_{1\, 7} -2 f_{1\, 8} +b_{4\, 10} & =0, 
 \\
2 f_{7\, 7} -2f_{7\, 8}  - b_{10\, 10} &=0. 
% \\ b_{1\, 4} -b_{4\, 7} +b_{4\, 10} &= 0, 
% \\ b_{1\, 1}-2b_{1\, 4} +b_{4\, 4} -b_{10\, 10}  &= 0, 
% \\  b_{1\, 1} +b_{1\, 10} - b_{4\, 4} +b_{4\, 10} &=0. 
% 
\end{align}
\end{subequations}

The number of linearly independent quadratic invariants under SO(3) is seven.
\end{thm}

An immediate corollary of Theorem \ref{thm1} is that the seven invariants  derived by  \citet{Ahmad02},  based partly on the work of \citet{Ting87}, are linearly independent and complete. 
The second  result is

\begin{thm}\label{thm2}
All quadratic invariants of $c$ under SO(2) are of the form $c^tFc$ where the 21$\times$21 symmetric matrix $F$ is 
\beq{4414}
F = F^{(1)} +F^{(2)},
\eeq
 $F^{(1)} $ and $F^{(2)}$ define  subspaces of dimension 29 and 6, respectively, and 
\beq{407}
 F^{(1)} = 
\begin{pmatrix}
A_{9\times 9} & 0_{9\times 12}  
\\ & \\
0_{12\times 9} & B^{(1)}_{12\times 12} \end{pmatrix} ,
\qquad
 F^{(2)} = 
\begin{pmatrix}
0_{9\times 9} & D_{9\times 12}  
\\ & \\ 
D^t_{12\times 9} & B^{(2)}_{12\times 12}  
\end{pmatrix} .
\eeq
\rev{The symmetric matrices $A $, $B^{(1)} $ and $B^{(2)} $ and the matrix $D$  are defined by 17, 27, 3 and 3 independent  elements, respectively}.  

With the axis of rotation  ${\bd e}_3$,  
\bal{415}
A &= \left(
\begin{tabular} {cccc cccc cccc}
       $f_{11}$ & $f_{12}$& $f_{13}$ & $f_{14}$ & $f_{15}$& $f_{16}$ &$f_{17}$     & $f_{18}$& $f_{19}$  
\\
&      $f_{11}$ & $f_{13}$& $f_{15}$ & $f_{14}$ & $f_{16}$& $f_{18}$ &$f_{17}$     & $f_{19}$ 
\\
& &    $f_{33}$ & $f_{34}$& $f_{34}$ & $f_{36}$ & $f_{37}$& $f_{37}$ &$f_{39}$      
\\
&&&    $f_{44}$ & $f_{45}$& $f_{46}$ & $f_{47}$ & $f_{48}$& $f_{49}$  
\\
&&&&   $f_{44}$ & $f_{46}$& $f_{48}$ & $f_{47}$ & $f_{49}$ 
\\
&&&&&  $f_{66}$ & $f_{67}$& $f_{67}$ & $f_{69}$  
\\
&S&Y&M&&&   $f_{77}$ & $f_{78}$& $f_{79}$  
\\
&&&&&&&  $f_{77}$ & $f_{79}$ 
\\
&&&&&&&& $f_{99}$  
\end{tabular} \right),
\\ & \nonumber \\
B^{(1)} &= \left(
\begin{tabular} {cccc cccc cccc}
       $b_{11}$ & 0& 0 & $b_{14}$ & 0& 0 &$b_{17}$     & 0& 0 & $b_{1\, 10}$     & 0 & 0 
\\
&      $b_{11}$ & 0& 0 & $b_{17}$ & 0& 0 &$b_{14}$     & 0& 0 & $b_{1\, 10}$ & 0 
\\
& &    $b_{33}$ & 0& 0 & $b_{36}$ & 0& 0 &$b_{36}$     & 0& 0 & $b_{3\, 12}$  
\\
&&&    $b_{44}$ & 0& 0 & $b_{47}$ & 0& 0 &$b_{4\, 10}$ & 0 & 0 
\\
&&&&   $b_{55}$ & 0& 0 & $b_{47}$ & 0& 0 &$b_{5\, 11}$ & 0 
\\
&&&&&  $b_{66}$ & 0& 0 & $b_{69}$ & 0& 0 &$b_{6\, 12}$  
\\
&&&&&&   $b_{55}$ & 0& 0 & $b_{5\, 11}$ & 0& 0 
\\
&&&&&&&  $b_{44}$ & 0& 0 & $b_{4\, 10}$ & 0 
\\
&&&&&&&& $b_{66}$ & 0& 0 & $b_{6\, 12}$  
\\
&S&Y&M&&&&&& $b_{10\, 10}$ & 0& 0  
\\
&&&&&&&&&&$b_{10\, 10}$ & 0   
\\
&&&&&&&&&&& $b_{12\, 12}$  
\end{tabular} \right). \label{4151}
\end{align}
The elements $f_{3\, 3}$, $f_{3\, 4}$, $f_{3\, 7}$ of $A$ and 
$b_{4\, 4}$ of $B^{(1)}$ are arbitrary, while  the 
remaining  40 distinct elements in $A$ and $B^{(1)}$  are arbitrary
 as long as they satisfy the 15 linearly independent conditions
\begin{subequations}\label{4021}
\bal{4021a}
 2f_{1\, 1}+2f_{1\, 2} - 4f_{1\, 6}  + f_{6\, 6} -f_{9\, 9}  &= 0,  
 \\
2f_{1\, 1}+2f_{1\, 2} -2f_{1\, 9} - f_{6\, 6}  -f_{6\, 9} &= 0,  
 \\
2f_{1\, 3}  - f_{3\, 6} -f_{3\, 9} &= 0,   
 \\
f_{1\, 4}+f_{1\, 5}  - f_{4\, 6} -f_{4\, 9} &= 0,    
 \\
f_{1\, 7}+f_{1\, 8} -f_{6\, 7} -f_{7\, 9}  &= 0,   
 \\ 
2f_{4\, 4} -2f_{4\, 5}  - b_{3\, 3} &= 0,  
 \\
2f_{1\, 4} - 2f_{1\, 5} + b_{3\, 6} &=0,  
 \\
2f_{4\, 7} - 2f_{4\, 8} - b_{3\, 12} &= 0,  
 \\
4f_{1\, 1} -2f_{1\, 6} -2f_{1\, 9} - b_{6\,6} & = 0,  
 \\
4f_{1\, 2} -2f_{1\, 6} -2f_{1\, 9} + b_{6\,9} &= 0,   
 \\
2f_{1\, 7} -2f_{1\, 8} + b_{6\, 12} &= 0,  
 \\
2f_{7\, 7} -2 f_{7\, 8} - b_{12\, 12} &= 0,  
 \\
b_{1\, 4} - b_{4\, 7} + b_{4\, 10} &= 0,  
 \\
   b_{1\, 1} -2b_{1\, 7} + b_{5\, 5}  - b_{10\, 10}  &= 0,    
 \\
 b_{1\, 1} +b_{1\, 10} - b_{5\, 5} +b_{5\, 11}  &= 0.  
\end{align}
\end{subequations}
Also, 
\bal{645}
B^{(2)} = \left(
\begin{tabular} {cccc cccc cccc}
       0 & 0 & 0 & 0 & $ b_{15}$& 0 & 0     & $ b_{18}$& 0 & 0    & $ b_{15}$ &  0
\\
&      0 & 0 & $- b_{18}$ & 0 & 0 &$- b_{15}$     & 0 & 0 & $ -b_{1\, 5}$     & 0 &  0 
\\
& &    0 & 0& 0 & 0 & 0& 0 & 0     & 0& 0 & 0 
\\
&&&    0 & $ b_{45}$& 0 &0     & 0 & 0 & 0    & $ b_{1, 8}+b_{4, 5}$ &  0
\\
&&&&   0& 0 & 0     & 0& 0 & $- b_{15}$     & 0 &  0 
\\
&&&&&  0 & 0& 0 & 0 & 0& 0 &0  
\\
&&&&&&  0     & $- b_{45}$& 0 & 0     & $ b_{15}$ &  0
\\
&&&&&&&  0& 0 & $- b_{1, 8}-b_{4, 5}$     & 0 &  0
\\
&&&&&&&& 0 & 0& 0 &0 
\\
&S&Y&M&&&&&& 0     & 0 &  0 
\\
&&&&&&&&&&0 & 0
\\
&&&&&&&&&&& 0
\end{tabular} \right),
\end{align}
and 
\bal{445}
D = \left(
\begin{tabular} {cccc cccc cccc}
0 & 0 & $d_{13}$ & 0 & 0 & 0 & 0 & 0 & 0 & 0 & 0 & $d_{1\, 12}$ 
\\ 
0 & 0 & $-d_{13}$ & 0 & 0 & 0 & 0 & 0 & 0 & 0 & 0 & $-d_{1\, 12}$ 
\\
0 & 0 & 0 & 0 & 0 & 0 & 0 & 0 & 0& 0 & 0 & 0 
\\ 
0 & 0 & 0 & 0 & 0 & $d_{13}$ & 0 & 0 & $d_{13}$ & 0 & 0 & $d_{4\, 12}$ 
\\
0 & 0 & 0 & 0 & 0 & $-d_{13}$ & 0 & 0 & $-d_{13}$ & 0 & 0 & $-d_{4\, 12}$ 
\\ 
0 & 0 & 0 & 0 & 0 & 0 & 0 & 0 & 0 & 0 & 0 & 0
\\
0 & 0 & $-d_{4\, 12}$ & 0 & 0 & $d_{1\, 12}$ & 0 & 0 & $d_{1\, 12}$ & 0 & 0 & 0
\\ 
0 & 0 & $d_{4\, 12}$ & 0 & 0 & $-d_{1\, 12}$ & 0 & 0 & $-d_{1\, 12}$ & 0 & 0 & 0 
\\
0 & 0 & 0 & 0 & 0 & 0 & 0 & 0 & 0 & 0 & 0 & 0
\end{tabular} \right), 
\end{align}
where the 6 elements $b_{15},\, b_{18},\, b_{45},\, d_{13},\,d_{1,\, 12},\,  d_{4,\, 12}$ are arbitrary. 

The number of linearly independent quadratic invariants under SO(2) is thirty five.

\end{thm}

\section{Invariants under reflection about a plane}\label{sec3}

\subsection{Linear and quadratic invariants under reflection}

In order to fix ideas  consider a tensor of order one,  a vector ${\bd v} = v_i {\bd e}_i$.  Invariants are defined in relation to a transformation, usually associated with some material symmetry. The simplest transformation is that of reflection about a plane orthogonal to a direction ${\bd e}$, which we denote $R({\bd e})$.  This transformation is associated with monoclinic symmetry \citep{cowin95}.  The action of $R({\bd e})$ on 
${\bd v} $ is defined by
\beq{-9}
 R( {\bd e}){\bd v} = {\bd v}- 2( {\bd v}\cdot{\bd e}) {\bd e} . 
 \eeq
Thus all ${\bd v}$ orthogonal to ${\bd e}$ are unchanged, or invariant under $R( {\bd e})$.  The  linear invariant of an arbitrary vector may be  {defined} as its projection onto the invariant subspace, i.e. ${\bd v} -( {\bd v}\cdot{\bd e}) {\bd e}$.  The {number} of linear invariants is the dimension of the invariant subspace, in this case 2.   

Quadratic invariants are defined by quadratic forms, which for a vector  require a symmetric second order tensor, say ${\bd F} = {\bd F}^t$.  Let
\beq{-8}
\phi ({\bd v})= {\bd v}\cdot {\bd F}\cdot{\bd v}, 
\eeq
then we seek ${\bd F}$ which leave $\phi$ unchanged under the action of $R( {\bd e})$ on ${\bd v}$: 
\beq{-23}
\phi =  R( {\bd e}) \phi \quad \text{where }  R( {\bd e}) \phi \equiv  \phi ( R( {\bd e}) {\bd v}).
\eeq
  Since 
\beq{-7}
 R( {\bd e})\phi = [{\bd v}- 2( {\bd v}\cdot{\bd e}) {\bd e}]\cdot {\bd F} \cdot[{\bd v}- 2( {\bd v}\cdot{\bd e}) {\bd e}]
 =  \phi - 4( {\bd v}\cdot{\bd e}) 
 \big[   {\bd v} - ( {\bd v}\cdot{\bd e}) {\bd e}\big] \cdot {\bd F}\cdot{\bd e},
 % \big[   {\bd e}\cdot {\bd F}\cdot{\bd v} - ( {\bd v}\cdot{\bd e}) {\bd e}\cdot {\bd F}\cdot{\bd e} \big], 
 \eeq
 we see that 
$ R( {\bd e})\phi = \phi$ under two circumstances: (i) for all ${\bd F}$ such that ${\bd F}\cdot{\bd e} = 0$, which defines a three dimensional subspace of second order symmetric tensors;  (ii) for all ${\bd F}$ of the form 
${\bd F} = \mu {\bd e} \otimes {\bd e}$, a one dimensional subspace.  It may be checked that all quadratic forms that leave $\phi$ fixed must be combinations of these, and  hence the number of quadratic invariants for vectors is 3+1=4.

As we deal with tensors of higher order it is simpler to work with the components relative to the basis ${\bd e}_i, i=1,2,3$.  To be specific, we consider ${\bd e} = {\bd e}_3$, then  the 2 linear invariants  are $v_1$ and $v_2$.  Similarly, the 4 quadratic  invariants are defined by  $\phi$ of the form \rf{-8} where $f_{13}=f_{23}=0$ but ${\bd F} $ is otherwise arbitrary.  In summary, 
\beq{-6}
{\bd v} = 
\begin{pmatrix} 
v_1\\
v_2  \\
0   
\end{pmatrix} ,\qquad
{\bd F} = 
\begin{pmatrix} 
f_{11} & f_{12} & 0\\
f_{12} & f_{22} & 0\\
0 & 0 & f_{33} 
\end{pmatrix}, \qquad n=1 \text{ under } R({\bd e}_3), 
\eeq
where $n=1$ indicates that these are the linear and quadratic invariants for   tensors of order $n$. 

Note that we do not present the quadratic invariants as, for instance, the set $\{ v_1^2,  v_2^2, v_1v_2,v_3^2\}$, but use notions from linear algebra which are natural for quadratic forms.

\subsubsection{n=2}

A second order symmetric tensor defines a unique quadratic form, through eq. \rf{-8}.  Based on the analysis for $n=1$ we can  immediately identify  the linear invariants under $R({\bd e}_3)$ any  second order symmetric tensor ${\bd V} = v_{ij}{\bd e}_i\otimes {\bd e}_j$ as  $v_{11}, v_{22}, v_{12}, v_{33}$.  

Quadratic forms on ${\bd V}$ are 
\beq{-11}
\phi = v_{ij}F_{ijkl}v_{kl}, 
\eeq
where the elements of $F$ satisfy 
\beq{-25}
F_{ijkl}= F_{jikl},\qquad
F_{ijkl}= F_{klij}.
\eeq
These are the same relations that define $\tens{E}$la, see eq. \rf{-2}. 
Thus, the symmetries of $F_{ijkl}$ under $R( {\bd e}_3)$ follow by standard arguments: all elements with index $3$ occurring an odd number of times must vanish \citep{cowin95}.  In summary, 
\beq{02}
{\bd V} = \begin{pmatrix} 
v_{11} & v_{12} & 0\\
v_{12} & v_{22} & 0\\
0 & 0 & v_{33} 
\end{pmatrix} ,\qquad
F = 
\begin{pmatrix} 
f_{11} & f_{12} & f_{13} & 0 & 0 & f_{16 }\\
f_{12} & f_{22} & f_{23} & 0 & 0 & f_{16 }\\
f_{13} & f_{23} & f_{33} & 0 & 0 & f_{36 }\\
0 & 0 &  0 & f_{44} & 0 & f_{46} \\
0 & 0 &  0 & 0 & f_{55} & 0 \\
f_{16 } & f_{26 } & f_{36 } &  0 & 0 & f_{66}  
\end{pmatrix},
\qquad n=2 \text{ under } R( {\bd e}_3). 
\eeq

The thirteen elements of $F$ define the subspace of $\tens{E}$la invariant under the single reflection $R( {\bd e}_3)$. That is, there are 13 linear invariants for fourth order symmetric elasticity tensors.  Since these tensors are defined by the quadratic form (energy) acting on second order symmetric tensors (strain) the 13 linear invariants of $n=4$ correspond to the quadratic invariants of $n=2$, in the same way that the linear invariants for $n=2$ correspond to the quadratic invariants of $n=1$, eqs. $\rf{-6}_2$ and $\rf{02}_1$.  It should be clear that the same equivalence holds for arbitrary tensor order $n$, and the result may be stated as follows:

\begin{lem}\label{lem1}
The quadratic invariants of a tensor of order $n$ are defined by the linear invariants of  the corresponding symmetric tensor of order $2n$.  
\end{lem}

\subsection{Remarks on reflections and rotations}

We are concerned with invariants under the group of proper orthogonal transformations, SO(3), and the group of rotations about an axis, SO(2).  The orthogonality requires that right-handed triads of coordinate axes remain right-handed under transformation, or equivalently that the determinants of the $3\times 3$ transformation matrices are $+1$.  But three dimensional inversion  is also acceptable by virtue of the fact that the tensor we are concerned  with is of even order, viz. 8.   Therefore, $SO(3)$  should be considered extended, to include the group $I_3$ of transformations $1,2,3\rightarrow -1,2,3$, etc.   Thus  $O(3) = SO(3)\cup I_3$.   
Similarly, $O(2) = SO(2)\cup I_2$ where $I_2$ is the group of transformations formed from  $1,2\rightarrow -1,2;\, 1,2\rightarrow 1,-2$.
  Another way to understand why we can replace $SO(2) \rightarrow O(2)$  is to consider  reflection about a plane, say $R({\bd e}_3)$,  followed by inversion.  This is equal to rotation about the normal to the plane by $\pi$, and hence $O(2)$ is equivalent to SO(2)$\cup R({\bd e}_3)$.  We will find this particularly useful as a starting point later, and it motivates consideration of the invariants under the action of $R({\bd e}_3)$ first.

We are now ready to determine the quadratic invariants of $\tens{E}$la, starting with invariants under reflection.

\section{Invariants of elastic moduli under reflection }\label{sec4}
%Wang and Zhao \cite{WangZhao03} obtained 20 invariants. 

We examine  $\tens{E}$la under reflection first about a single plane, and then about two.  
Based on the Voigt notation for indexing elastic moduli with 21 elements, the indices for $F$ of \rf{-1} and \rf{-3} as a square symmetric matrix $f_{ij}$ run from $1,2,\ldots , 21$, according to the following

\begin{tabular}{r ccccccc ccccccc ccccccc}
i =& 1 & 2 &3 &4 &5 & 6 &7 &8 &9 &10 &11 &12 &13 &14 &15 &16 &17 &18 &19 &20 &21 
  \\ 
IJ =& 11 &22 &33 &23 &13 &12 & 44 &55 &66 &14 &25 &36 &34 &15 &26 &24 &35 &16 &56 &46 &45  
\end{tabular}
where $IJ$ are the Voigt indices, 
which are  capital suffices taking the values $1,2,\ldots, 6$ by
\beq{904}
I=1,2,3,4,5, 6 \qquad \Leftrightarrow \qquad  ij = 11,22,33,23,31,12 .  
\eeq
For example, $f_{13\, 17} =f_{34 35} = f_{33233313}$ using the two, four and eight index notations. In this way the elements $f_{ijklpqrs}$ are represented uniquely by the  $\frac12(21\times22)=231$ elements 
$f_{ij}$.

Consider first the reflection $R( {\bd e}_3)$ about the    $x_1x_2$ plane as a symmetry plane.   Since we are dealing with a tensor of even order (8),  it follows that  those  elements vanish that have indices $ijklpqrs$ where $3$ occurs an \emph{odd} number of times. 
There are $13\times 8 = 104$  elements in this category: 
\beq{401}
f_{j_3 k_3}=0, \qquad j_3 \in \{1,\ldots , 9,12,15, 18, 21 \}, \quad k_3 \in \{10,11,13,14,16,17,19,20 \}.
\eeq
That leaves $231-104=127$ non-zero elements that remain under reflection about a symmetry plane, or monoclinic symmetry. 

Next consider reflection about two planes with perpendicular normals, say $R( {\bd e}_3)$ and $R( {\bd e}_1)$.  As before, 
elements  that have indices $ijklpqrs$ where $1$ occurs an \emph{odd} number of times must vanish. 
The $104$ elements with $1$ occurring  an {odd} number of times are
\beq{411}
f_{j_1 k_1}=0, \qquad j_1 \in \{ {\bf 1,\ldots , 9},10, 13, 16,19 \}, \quad k_1 \in \{{\bf 11},12,{\bf 14},15,{\bf 17},18,{\bf 20},21 \}.
\eeq
But some of these coincide with members of \rf{401}: the elements indicated by the indices in boldface, plus the elements defined by the remaining indices.  The   elements that  are distinct from \rf{401} are the $9\times 4 = 36$: 
\beq{412}
f_{\bar{j}_1 \bar{k}_1}=0, \qquad \bar{j}_1 \in \{  1,\ldots , 9 \}, \quad \bar{k}_1 \in \{
12,15,18,21 \}, 
\eeq
and the $4\times 4 = 16$:
\beq{413}
f_{\hat{j}_1 \hat{k}_1}=0, \qquad \hat{j}_1 \in \{ 10, 13, 16, 19 \}, \quad \hat{k}_1 \in \{
 11, 14, 17, 20 \}, 
\eeq
In summary, $F$ is of the form 
\beq{414}
F = \begin{pmatrix}
A_{9\times 9} & 0_{9\times 12}  \\
0_{12\times 9} & B_{12\times 12} 
\end{pmatrix},
\eeq
where $A$ and $B$ are symmetric.  The matrix $A$ is full, indicating $9\times 10/2 = 45$ independent elements, and \emph{letting the indices of $B$ run from $1$ to $12$}, we have
\beq{45}
b_{ij}=0, \qquad (ij) \in \big\{ \{(3,6,9,12)\times(1,2,4,5,7,8,10,11)\}\oplus \{(1,4,7,10)\times(2,5,8,11)\}
\big\}.
\eeq
Therefore, $B$ has $12\times13/2 -4\times 12 = 30$  for a total of $75$ independent elements in $F$.  In fact, $B$ has banded structure 
\bal{0456}
B = \left(
\begin{tabular} {cccc cccc cccc}
       $b_{11}$ & 0& 0 & $b_{14}$ & 0& 0 &$b_{17}$     & 0& 0 & $b_{1\, 10}$     & 0 & 0 
\\
&      $b_{22}$ & 0& 0 & $b_{25}$ & 0& 0 &$b_{28}$     & 0& 0 & $b_{2\, 11}$ & 0 
\\
& &    $b_{33}$ & 0& 0 & $b_{36}$ & 0& 0 &$b_{39}$     & 0& 0 & $b_{3\, 12}$  
\\
&&&    $b_{44}$ & 0& 0 & $b_{47}$ & 0& 0 &$b_{4\, 10}$ & 0 & 0 
\\
&&&&   $b_{55}$ & 0& 0 & $b_{58}$ & 0& 0 &$b_{5\, 11}$ & 0 
\\
&&&&&  $b_{66}$ & 0& 0 & $b_{69}$ & 0& 0 &$b_{6\, 12}$  
\\
&&&&&&   $b_{77}$ & 0& 0 & $b_{7\, 10}$ & 0& 0 
\\
&&&&&&&  $b_{88}$ & 0& 0 & $b_{8\, 11}$ & 0 
\\
&&&&&&&& $b_{99}$ & 0& 0 & $b_{9\, 12}$  
\\
&&&&&&&&& $b_{10\, 10}$ & 0& 0  
\\
&&&&&&&&&&$b_{11\, 11}$ & 0   
\\
&&&&&&&&&&& $b_{12\, 12}$  
\end{tabular} \right). 
\end{align}
Note the  indices for $B$ as a square symmetric matrix $b_{ij}$ run from $1,2,\ldots , 12$, according to the following

\begin{tabular}{r ccccccc ccccccc ccccccc}
B & & i &=& 1 & 2 &3 &4 &5 & 6 &7 &8 &9 &10 &11 &12 
  \\ 
Voigt &  & IJ& =&14 &25 &36 &34 &15 &26 &24 &35 &16 &56 &46 &45  
\\ 
F(21$\times$21) &  & i &=&10 &11 &12 &13 &14 &15 &16 &17 &18 &19 &20 &21 
\end{tabular}

The 8$^{th}$ order tensor $F$ of eqs. \rf{414} and \rf{0456} is unchanged under reflection about a third  plane orthogonal to the others, i.e. $R( {\bd e}_2)$.  In order to see this, note that $R( {\bd e}_2)$ implies that all elements with index 2 occurring an odd number of times must vanish.  However, since the tensor is of even order, and we have eliminated elements with 1 and 3 occurring odd numbers of times, the tensor is  automatically invariant under $R( {\bd e}_2)$.  This effect  is well known in elasticity: that a material with two orthogonal planes of symmetry automatically has a third \citep{cowin95}. 

In summary, 

\begin{lem}\label{lem2}
All quadratic invariants of $c$ under reflection about a plane are of the form $c^tFc$ where the 21$\times$21 symmetric matrix $F$ is
\beq{0414}
F = \begin{pmatrix}
A_{9\times 9} & D_{9\times 12}  \\
D^t_{12\times 9} & B_{12\times 12} 
\end{pmatrix},
\eeq
  $A=A^t$ is full  and  $B=B^t$ and $D$ are of banded  form,
\bal{1456}
B = \left(
\begin{tabular} {cccc cccc cccc}
       $b_{11}$ & $\bf b_{12}$ & 0 & $b_{14}$ & $\bf b_{15}$& 0 &$b_{17}$     & $\bf b_{18}$& 0 & $b_{1\, 10}$     & $\bf b_{1\, 11}$ &  0
\\
&      $b_{22}$ & 0 & $\bf b_{24}$ & $b_{25}$& 0 &$\bf b_{27}$     & $b_{28}$& 0 & $\bf b_{2\, 10}$     & $b_{2\, 11}$ &  0 
\\
& &    $b_{33}$ & 0& 0 & $b_{36}$ & 0& 0 &$b_{39}$     & 0& 0 & $b_{3\, 12}$  
\\
&&&    $b_{44}$ & $\bf b_{45}$& 0 &$b_{47}$     & $\bf b_{48}$& 0 & $b_{4\, 10}$     & $\bf b_{4\, 11}$ &  0
\\
&&&&   $b_{55}$& 0 &$\bf b_{57}$     & $b_{58}$& 0 & $\bf b_{5\, 10}$     & $b_{5\, 11}$ &  0 
\\
&&&&&  $b_{66}$ & 0& 0 & $b_{69}$ & 0& 0 &$b_{6\, 12}$  
\\
&&&&&&   $b_{77}$     & $\bf b_{78}$& 0 & $b_{7\, 10}$     & $\bf b_{7\, 11}$ &  0
\\
&&&&&&&  $b_{88}$& 0 & $\bf b_{8\, 10}$     & $b_{8\, 11}$ &  0
\\
&&&&&&&& $b_{99}$ & 0& 0 & $b_{9\, 12}$  
\\
&&&&&&&&& $b_{10\, 10}$     & $\bf b_{10\, 11}$ &  0 
\\
&&&&&&&&&&$b_{11\, 11}$ & 0
\\
&&&&&&&&&&& $b_{12\, 12}$  
\end{tabular} \right)
\end{align}
and
\bal{045}
D = \left(
\begin{tabular} {cccc cccc cccc}
0 & 0 & $d_{13}$ & 0 & 0 & $d_{16}$ & 0 & 0 & $d_{19}$ & 0 & 0 & $d_{1\, 12}$ 
\\ 
0 & 0 & $d_{23}$ & 0 & 0 & $d_{26}$ & 0 & 0 & $d_{29}$ & 0 & 0 & $d_{2\, 12}$ 
\\
0 & 0 & $d_{33}$ & 0 & 0 & $d_{36}$ & 0 & 0 & $d_{39}$ & 0 & 0 & $d_{3\, 12}$ 
\\ 
0 & 0 & $d_{43}$ & 0 & 0 & $d_{46}$ & 0 & 0 & $d_{49}$ & 0 & 0 & $d_{4\, 12}$ 
\\
0 & 0 & $d_{53}$ & 0 & 0 & $d_{56}$ & 0 & 0 & $d_{59}$ & 0 & 0 & $d_{5\, 12}$ 
\\ 
0 & 0 & $d_{63}$ & 0 & 0 & $d_{66}$ & 0 & 0 & $d_{69}$ & 0 & 0 & $d_{6\, 12}$ 
\\
0 & 0 & $d_{73}$ & 0 & 0 & $d_{76}$ & 0 & 0 & $d_{79}$ & 0 & 0 & $d_{7\, 12}$ 
\\ 
0 & 0 & $d_{83}$ & 0 & 0 & $d_{86}$ & 0 & 0 & $d_{89}$ & 0 & 0 & $d_{8\, 12}$ 
\\
0 & 0 & $d_{93}$ & 0 & 0 & $d_{96}$ & 0 & 0 & $d_{99}$ & 0 & 0 & $d_{9\, 12}$ 
\end{tabular} \right), 
\end{align}
where the second index of $D$ is the same as that for $B$, i.e. runs from $1,2,\ldots , 12$.
The number of linearly independent quadratic invariants is 127.  

Under reflection  about two orthogonal planes  the number of linearly independent quadratic invariants reduces to 75, and $F$ then has the form 
\rf{0414} with $D\equiv 0$ and the  16 elements in bold in \rf{1456}  are zero, i.e. $B$ reduces to $B^{(1)}$ of eq. \rf{4151}. 

\end{lem}

Thus, invariance under reflection about one and then two orthogonal planes reduces the number of independent elements in $F$ from 231 to 127 to 75.  The case of a single plane is monoclinic symmetry, and two (or three) corresponds to orthorhombic symmetry.   

Reflection symmetry is an  important tool in  developing the structure of anisotropic elastic tensors \citep{cowin95}. In fact, all eight fundamental symmetries can be cast in terms of repeated application of the reflection operator \citep{CVC}. 
The  definition of transverse isotropy implies that every plane containing the axis of symmetry is a symmetry plane, in addition to the plane orthogonal to the axis.   However, this equivalence \emph{does not} hold in the present problem, dealing with the form of the 8$^{th}$ order tensor $F$ invariant under SO(2).  Consider rotation about ${\bd e}_3$ by $\pi$, which transforms $\{ {\bd e}_1, {\bd e}_2\} \rightarrow \{ {\bd e}_2, -{\bd e}_1\} $.  Since elements with odd numbered occurrences of $3$ have already been eliminated by the requirement of invariance under  $R({\bd e}_3)$, it therefore eliminates terms with 1 (or 2) occurring an odd number of times.  Thus, $c_{3312}=0$ for TI elasticity, but it does not eliminate elements such as $f_{33122233}$ from $F$.  Hence SO(2) invariance for $F$ is not the same as invariance under $R({\bd e})$ for all ${\bd e} \perp {\bd e}_3$. 

In other words, SO(2) invariance of $F$ is  not  a subspace of orthorhombic symmetry, 
but it is a subspace of monoclinic symmetry.  This allows us to start the search for  
SO(2) invariance with only 127 of the 231 elements.  Furthermore, SO(3) invariance of $F$ is    a proper subspace of orthorhombic symmetry, meaning it can be sought starting from the  75 element form of $F$ in \rf{33}.   

First we need to introduce the rotation matrix for $\tens{E}$la.

\section{Invariants under rotation about an axis }\label{sec5}

We commence by casting the problem in general terms, with some simpler examples before considering the problem for the 8$^{th}$ order tensor. 

\subsection{General theory}

Let $c$ be a vector, not restricted to the elastic modulus vector, and consider the quadratic form 
\beq{1}
\phi =  c^tFc,
\eeq
where $F$ is symmetric,\footnote{If $F=M+S$ where $M=M^t$ and $S=-S^t$  then 
$\phi$ defined in \rf{1} is $\phi =  c^tFc$ independent of $S$.  The skew-symmetric part of $F$ is irrelevant and can be set to zero.}
\beq{2}
F=F^t . 
\eeq
Under rotation through angle $\theta$ about axis ${\bf p}$, a 3-vector, 
\beq{3}
c \rightarrow c' = Qc, 
\eeq
where $Q= Q(\theta, {\bf p}) $ is the rotation. Accordingly, 
\beq{4}
\phi \rightarrow \phi' (\theta)=   c^tQ^tFQc .  
\eeq
We want $\phi' (\theta)$ to be independent of $\theta$, that is, its derivative should vanish for all $\theta$.  The derivative is 
\beq{5}
\dot{\phi}' =   c^tQ^tF\dot{Q}c +   c^t\dot{Q}^tFQc    ,
\eeq
where the dot indicates differentiation with respect to $\theta$. 
By  an application of standard Lie group and  Lie algebra
theory \citep{Fegan},   $Q(\theta)$  is a one parameter subgroup of SO(N) where $N$ is the vector length.  $Q$ is generated by
$P = -P^t$, an element from its Lie algebra of skew-symmetric matrices.  Thus,  $Q(\theta) = \exp \theta P$, from which the derivative is 
\beq{6}
\dot{Q}  = PQ. 
\eeq
The particular form of $P$ for elasticity tensors $(N=21)$ was derived by \citet{Norris06}.

Equations \rf{5} and \rf{6} imply 
\beq{7}
\dot{\phi}' = c^t \big(   Q^tFPQ - Q^tPFQ \big)c . 
\eeq
Note that the matrix $(   Q^tFPQ - Q^tPFQ)$ is symmetric, hence $\dot{\phi}'(\theta)=0$  for all possible $c$  implies  the SO(2) invariance condition: 
\beq{9}
   Q^t (FP  -  PF)Q = 0 \quad \text{for all } \theta . 
\eeq
Hence we have 
\begin{lem}\label{lem2.5}
The quadratic form $\phi =  c^tFc$ is an SO(2) invariant if and only if $F$ commutes with the generator $P$: 
\beq{11}
 FP  -  PF = 0 . 
\eeq
\end{lem}
 Equation  \rf{11} is a necessary condition  since \rf{9} must hold at $\theta = 0$ where $Q(0) = I$.  The sufficiency of \rf{11} is clear, hence eq. \rf{11} is the condition that $F$ must satisfy.

The question of quadratic invariants under SO(2) therefore reduces to finding all symmetric $F$ which commute with the skew rotation generator $P$, or equivalently, $\sym (FP) $ vanishes.
We next demonstrate this approach by application to increasingly higher order tensors, associated with quadratic forms on tensors of  order $n=1,2$ and ultimately the desired $n=4$. 

\subsection{Quadratic forms on vectors: n=1}

In this case the generator is the standard and well known skew symmetric second order tensor defined by the axis of rotation ${\bd p}$, 
\beq{20}
P_{ij} = -\epsilon_{ijk}p_k . 
\eeq
Let ${\bf p}={\bf e}_3$ so that 
\beq{201}
P = \begin{pmatrix} 
0 & 1 & 0 \\ 
-1 & 0 & 0 \\ 
0 & 0 & 0
\end{pmatrix} ,
\eeq
and let 
\beq{2001}
F = \begin{pmatrix} 
f_{11} & f_{12} & f_{13} \\ 
f_{12} & f_{22} & f_{23} \\ 
f_{13} & f_{23} & f_{33} 
\end{pmatrix} .
\eeq
Then 
\beq{22}
FP-PF = \begin{pmatrix} 
-2f_{12} & f_{11}-f_{22} & -f_{23} \\ 
f_{11}-f_{22}& 2f_{12} & f_{13} \\ 
-f_{23} & f_{13} & 0 
\end{pmatrix} .
\eeq
Setting each of the elements to zero gives four conditions: 
$f_{11}-f_{22} = 
f_{12} = 
f_{13} = 
f_{23} = 0$,
so that the most general $F$ is of the form
\beq{24}
F = \begin{pmatrix} 
a &  0 & 0 \\ 
0& a & 0 \\ 
0 & 0 & b 
\end{pmatrix} ,
\eeq
for $a,b \ne 0$. 

The associated invariants are 
\beq{25}
\phi_a = c_1^2 + c_2^2, \qquad \phi_b = c_3^2. 
\eeq
For arbitrary direction ${\bf p}$ these are 
\beq{26}
\phi_a = c^tc - (p^tc)^2, \qquad \phi_b = (p^tc)^2,  
\eeq
corresponding to 
\beq{27}
F_a =  - P^2, \qquad F_b = I +P^2. 
\eeq
Any $F$  comprised of even powers of $P$ will obviously commute with  $P$.  For $n=1$ there are only two independent $F$ of this form, i.e. $F=I, P^2$, since $P$ satisfies the characteristic equation
\beq{28}
P^3+P = 0, 
\eeq
and hence $P^{2+2m} = (-1)^m P^2$. 

\subsection{Quadratic forms on tensors: n=2}

The relation between the second order symmetric tensor $C$ and the 6-vector $c$ is 
\beq{200}
C = \begin{pmatrix}
 C_{11}& C_{12}& C_{13} 
 \\
 C_{12}& C_{22}&C_{23}
 \\
 C_{13}&C_{23}&C_{33}
\end{pmatrix}
\qquad
\Rightarrow 
\qquad
c \equiv  %\begin{pmatrix}  c_1 \\ c_2 \\ c_3 \\ c_4 \\ c_5 \\ c_6 \end{pmatrix} = 
 \big( C_{11} \   C_{22} \  C_{33} \  \sqrt{2} C_{23} \  \sqrt{2} C_{13} \  \sqrt{2} C_{12} \big)^t. 
\eeq
$P$ is now a 6$\times$6 skew symmetric matrix, 
\beq{31}
P = 
\begin{pmatrix}
0 & \sqrt{2}(Y -Z^t)
       \\  &  \\
\sqrt{2}(Z -Y^t) & X -X^t
\end{pmatrix}
\eeq
where
\beq{32}
X = \begin{pmatrix}
0  & p_3  & 0 \\ % && \\ 
0 & 0& p_1 \\ % && \\ 
 p_2 & 0 & 0
\end{pmatrix} ,
\quad
Y = \begin{pmatrix} 
0  & p_2  & 0 \\ % && \\ 
0 & 0& p_3 \\ % && \\ 
 p_1 & 0 & 0
\end{pmatrix} ,
\quad
Z = \begin{pmatrix} 
0  & p_1  & 0 \\ % && \\ 
0 & 0& p_2 \\ % && \\ 
 p_3 & 0 & 0 
\end{pmatrix}  . 
\eeq
The matrix   $P$ was first derived by \citet{mcj}, and the present form is due to \cite{Norris06}. We will find this useful when we consider the analogous $P$ for $n=4$ later.  We note that the characteristic equation of $P$ is now of fifth (2n+1) order \citep{mcj,Norris06},
\beq{281}
P(P^2+I)(P^2+4I) = 0. 
\eeq

Let ${\bf p}={\bf e}_3$ then
\beq{33}
P  =  \begin{pmatrix} 
0 & 0 & 0 & 
        0 &   0 & -\sqrt{2}   
\\ % & & & & & \\
0 & 0 & 0 & 
         0 &0 &  \sqrt{2}  
\\ % & & & & & \\
0 & 0 & 0 & 
          0 &    0 &0
\\ % & & & & & \\
0 &   0 &   0   & 0 & 1 &  0
\\ % & & & & & \\
  0 &0 &   0  & -1 & 0 & 0
\\ % & & & & & \\
\sqrt{2}   &  -\sqrt{2}   &0 &0 & 0 & 0 
\end{pmatrix} , 
\eeq
and the SO(2) commutator is 
\beq{34}
  \frac{FP-PF}{ \sqrt{2} }  = 
%\nonumber \\
  \begin{pmatrix} 
-2f_{16} & f_{16}-f_{26} & -f_{36} & \frac1{\sqrt{2}}f_{15}-f_{46} & -\frac1{\sqrt{2}}f_{14}-f_{56} & f_{11}-f_{12}- f_{66} 
  \\ & & & & & \\
 & 2f_{26} & f_{36} & \frac1{\sqrt{2}}f_{25}+f_{46} & -\frac1{\sqrt{2}}f_{24}+f_{56} & f_{12}-f_{22}+f_{66}
 \\ & & & & & \\
  &   & 0 & \frac1{\sqrt{2}}f_{35}  & -\frac1{\sqrt{2}}f_{34}  & f_{13}-f_{23} 
\\ & & & & & \\
  &   &   &  {\sqrt{2}}f_{45} & 	 \frac1{\sqrt{2}}(f_{55}-f_{44}) & f_{14}-f_{24} + \frac1{\sqrt{2}}f_{56}
\\ & & & & & \\
    S   &  Y  & M & & -{\sqrt{2}}f_{45} & f_{15}-f_{25} -\frac1{\sqrt{2}}f_{46}
\\ & & & & & \\
   &   &  &   &   & 2f_{16}-2f_{26}
\end{pmatrix} .
\eeq% nd{align}
Setting all elements to zero  and solving the resulting linear equations, we find that $F$  has five independent elements (and the expected form of a TI  tensor of elastic moduli),
\beq{35}
F = \begin{pmatrix} 
f_{11} & f_{12} & f_{13}  &0 &0 &0 
\\
f_{12} & f_{11} & f_{13}  &0 &0 &0 
\\ 
f_{13} & f_{13} & f_{33}  &0 &0 &0 
\\
0 &0 &0 & f_{44} &0 &0 
\\
0 &0 &0 &0 &f_{44} &0 
\\
0&0 &0 &0 &0 & f_{11}-f_{12}
\end{pmatrix} .
\eeq
Thus, there are five  quadratic invariants,  
\beq{36}
\begin{matrix}
 \phi_{11} = C_{11}^2+C_{22}^2+2C_{12}^2, &
   \quad \phi_{12}   = C_{11} C_{22}-C_{12}^2, &
    \quad  \phi_{33} = C_{33}^2 ,
   \\  
\phi_{13} = (C_{11} +C_{22})C_{33} , &
 \quad \phi_{44}  = C_{13}^2+C_{23}^2, &
 \end{matrix}
  \eeq
corresponding to $f_{11}$,  $f_{12}$,  $f_{33}$,  $f_{13}$,  and $f_{44}$, respectively.

The linear invariants of $C$ under SO(2) are 
\beq{37}
\lambda_1 = C_{11} +C_{22}, \qquad \lambda_2 = C_{33}.
\eeq
The five quadratic invariants can be recast as three defined by the linear invariants and two new ones, e.g., 
\beq{38}
 \phi_1=\lambda_1^2, 
 \quad
   \phi_2=\lambda_2^2, 
 \quad
  \phi_3=\lambda_1 \lambda_2,   
 \quad
 \phi_4 = C_{11} C_{22}-C_{12}^2,
  \quad
 \phi_5 = C_{13}^2+C_{23}^2.
\eeq

In keeping with the statement of Lemma \ref{lem1}, we note that the five independent elements of $F$ correspond to the linear invariants of $\tens{E}$la under SO(2).  We next consider the quadratic invariants of $\tens{E}$la under SO(2).

\subsection{Quadratic forms on the elastic moduli: n=4}

The fundamental quadratic form is now expressed in alternative ways,
\beq{101}
\phi = c^t F c =  \tilde{c}^t \tilde{F} \tilde{c} , 
\eeq
where $\tilde{c}$ is a 21-vector related to the 21-vector of  moduli $c$ in \rf{00} by 
\beq{132} 
\tilde{c} = Tc, 
\eeq
and 
\beq{2054}
T = \diag \big(1\ \,1\ \,1 \ \,\sqrt{2}\ \,\sqrt{2}\ \,\sqrt{2}\ \,2\ \,2\ \,2\ \,2\ \,2\ \,2\ \,2\ \,2\ \,2\ \,2\ \,2\ \,2\ \,
\sqrt{8}\ \,\sqrt{8}\ \,\sqrt{8}\big).
\eeq
We  introduce $\tilde{F} $ in \rf{101} to remove the $\sqrt{2}$ factors from the final expressions. The two are simply  related by 
\beq{2051}
F = T \tilde{F} T. 
\eeq
Results below are given for the elements of $F$ which  do not have the $\sqrt{2}$ factors.

\subsubsection{The rotation matrix}

The 21$\times$21 skew symmetric generator is  \citep{Norris06}
\beq{41}
P = R-R^t,
\eeq
where 
\beq{42}
R = 
  \begin{pmatrix} 
{  0} & {  0} & {  0} & {  0} &2{  Y}  & {  0} &{  0} 
					\\ % && && && \\
{  0} & {  0} & {  0} & -\sqrt{2}{  Y}  & {  0}  & \sqrt{2}{  N}  &{  0} 
					\\ % && && && \\
{  0} & {  0} & {  0} & {  0} & {  0} & 2{  N}  & -\sqrt{2}{  Y}  
					\\ % && && && \\
{  0} & -\sqrt{2}{  Z} & {  0} & {  0} & {  X} &{  0} &  -\sqrt{2}{  X}  
					\\ % && && && \\
{  0}  &\sqrt{2}{  N} & 2{  N} & {  0} & {  0} & {  X}  & {  0}
					\\ % && && && \\
 2{  Z}   & {  0} & {  0} &   {  X} & {  0} & {  0} &   \sqrt{2}{  X}
   				\\ % && && && \\
{  0} &{  0} & -\sqrt{2}{  Z}  & - \sqrt{2}{  X}   &\sqrt{2}{  X} & {  0} &  - {  X}
\end{pmatrix},
\eeq
with $X$, $Y$ and $Z$ as before in eq. \rf{32}, and 
\beq{43}
 N = \begin{pmatrix} 
p_1  & 0  & 0 \\ % && \\ 
0 & p_2& 0 \\ % && \\ 
0 & 0 & p_3
\end{pmatrix} .
\eeq
Note that the characteristic equations for $P$ is 
\beq{481}
P(P^2+I)(P^2+4I)(P^2+9I)(P^2+16I) = 0.   
\eeq

The conditions \rf{11} are that 
the commutator of $\tilde{F}$ and $P$ vanish, 
\beq{033}
\tilde{F}P-P\tilde{F} = 0. 
\eeq
This may be rewritten in a form involving the elements of $F$ of \rf{2051}, with the intent of removing the factors of $\sqrt{2}$.  The resulting equation is 
\beq{113}
\hat{P} F + F \hat{P}^t = 0, \qquad \text{where  } \hat{P} =  TPT^{-1}. 
\eeq
The modified generator $\hat{P}$ is no longer skew symmetric, but it still satisfies the characteristic equation \rf{481}, and, like $P$, can be used to compute the rotation $Q=\exp (\theta P)$, or its modified form, $\hat{Q}=\exp (\theta \hat{P}) = TQT^{-1}$, as a polynomial of degree 8 in $\hat{P}$ \citep{Norris06}. With ${\bd p} = {\bd e}_3$,  $R$ has 18 non-zero elements, hence $\hat{P}$ has $36$ non-zero elements. 

%$F$ has $21(21 +1)/2 = 231$ independent elements.  Recall that for $n=1$ and $2$ the number of independent elements of $F$ were $6$ and $15$, and we found $2$ and $5$ quadratic invariants, respectively.   That is, the number of quadratic invariants for both $n=1$ and $2$  is one third the number of independent elements of $F$.  If the same one-third rule holds for $n=4$ we would have $77$  quadratic invariants for the  elastic moduli.  

\subsubsection{Quarter turn conditions on the elements of F}

Before applying the SO(2) condition \rf{113} to $F$, we perform a preliminary simplification by invoking  invariance under a quarter turn about the axis of rotation.   The transformed elements $F_{ijklpqrs}$ should be unaltered under the interchanges of indices $1, 2\rightarrow 2, -1$.  This implies a total of 52 relations among the   127 elements of $F$, which we split into two categories depending on whether the index $1$ occurs an even or an odd number of times.  In the former category 
are the following 31 identities, 
\beq{402}
\begin{matrix}
f_{1\, 1}=f_{2\, 2}  ,& 
f_{1\, 3}=f_{2\, 3} , &
f_{1\, 4}=f_{2\, 5} , &
f_{1\, 5}= f_{2\, 4}  , 
\\ &&& \\
f_{1\, 6}=f_{2\, 6} , &
f_{1\, 7}=f_{2\, 8} , &
f_{1\, 8}=f_{2\, 7} , &
f_{1\, 9}=f_{2\, 9} , 
\\ &&& \\
f_{3\, 4}=f_{3\, 5} , &
f_{3\, 7}=f_{3\, 8} , & 
f_{4\, 4} =f_{5\, 5} , &
f_{4\, 6}= f_{5\, 6} ,
\\ &&& \\ 
f_{4\, 7}=f_{5\, 8} , & 
f_{4\, 8}=f_{5\, 7} , &
f_{4\, 9} = f_{5\, 9} , &
f_{6\, 7}=f_{6\, 8} , &
\\ &&& \\ 
f_{7\, 7}=f_{8\, 8} , &
f_{7\, 9}=f_{8\, 9} , & 
&
\\ &&& \\ 
b_{1\, 1}=b_{2\, 2} , &
b_{1\, 4} = b_{2\, 8}, &
b_{1\, 7} = b_{2\, 5}, &
b_{1\, 10} = b_{2\, 11}, 
\\ &&& \\ 
b_{3\, 6} = b_{3\, 9}, &
b_{4\, 4}=b_{8\, 8} , &
b_{4\, 7} = b_{5\, 8},  &
b_{4\, 10}= b_{8\, 11}, 
\\ &&& \\ 
b_{5\, 5}=b_{7\, 7} ,  &
b_{5\, 11} = b_{7\, 10}, &
b_{6\,6}  = b_{9\, 9}, &
b_{6\, 12}= b_{9\, 12},
\\ &&& \\ 
b_{10\, 10}=b_{11\, 11} , &
&& 
\end{matrix} 
\eeq
and in the second group we have the following 21 connections, 
\beq{1402}
\begin{matrix}
b_{1\, 5}+b_{2\, 7} =0  ,& 
b_{1\, 8}+ b_{2\, 4} =0, &
b_{1\, 11}+b_{2\, 10} =0,  
\\ &&& \\
b_{4\, 5}+b_{7\, 8} =0 ,& 
b_{4\, 11} +b_{8\, 10} =0, &
b_{5\, 10}+b_{7\, 11} =0,  
\\ &&& \\
d_{1\, 3} +d_{2\, 3} =0 ,& 
d_{4\, 3} +d_{5\, 3}=0  , &
d_{7\, 3} +d_{8\, 3}=0 ,  
\\ &&& \\
d_{1\, 6} +d_{2\, 9}=0 ,& 
d_{1\, 9} + d_{2\, 6}=0, &
d_{3\, 6} + d_{3\, 9}=0,
\\ &&& \\
d_{4\, 6} +d_{5\, 9}=0 ,& 
d_{4\, 9} + d_{5\, 6}=0, &
d_{6\, 6} + d_{6\, 9}=0,
\\ &&& \\
d_{7\, 6} +d_{8\, 9}=0 ,& 
d_{7\, 9} +d_{8\, 6}=0, &
d_{9\, 6} +d_{9\, 9}=0,
\\ &&& \\
d_{1\, 12}  +d_{2\, 12}=0,& 
d_{4\, 12} +d_{5\, 12}=0 , &
d_{7\, 12} +d_{8\, 12}=0. 
\end{matrix} 
\eeq

Therefore, $A$,  $B$ and $D$ have 45-18=27, 46-19=27 and 36-15=21 independent elements, respectively, for a total of 127- 52= 75 independent elements in $F$.  $A$ has the form given in \rf{415}, while $B$ and $D$ are of the form
\bal{-34}
B &= \left(
\begin{tabular} {cccc cccc cccc}
       $b_{11}$ & $b_{12}$ & 0 & $b_{14}$ & $b_{15}$& 0 &$b_{17}$     & $b_{18}$& 0 & $b_{1\, 10}$     & $b_{1\, 11}$ & 0 
\\
&      $b_{11}$ & 0& $-b_{18}$ & $b_{17}$ & 0& $-b_{18}$ &$b_{14}$     & 0& $-b_{1\, 11}$ & $b_{1\, 10}$ & 0 
\\
& &    $b_{33}$ & 0& 0 & $b_{36}$ & 0& 0 &$b_{36}$     & 0& 0 & $b_{3\, 12}$  
\\
&&&    $b_{44}$ & $b_{45}$ & 0 & $b_{47}$ & $b_{48}$& 0 &$b_{4\, 10}$ & $b_{4\, 11}$ & 0 
\\
&&&&   $b_{55}$ & 0& $b_{57}$ & $b_{47}$ & 0& $b_{5\, 10}$  &$b_{5\, 11}$ & 0 
\\
&&&&&  $b_{66}$ & 0& 0 & $b_{69}$ & 0& 0 &$b_{6\, 12}$  
\\
&&&&&&   $b_{55}$ & $-b_{45}$ & 0 & $b_{5\, 11}$ & $-b_{5\, 10}$& 0 
\\
&&&&&&&  $b_{44}$ & 0& $-b_{4\, 11}$ & $b_{4\, 10}$ & 0 
\\
&&&&&&&& $b_{66}$ & 0& 0 & $b_{6\, 12}$  
\\
&&&&&&&&& $b_{10\, 10}$ & $b_{10\, 11}$& 0  
\\
&&&&&&&&&&$b_{10\, 10}$ & 0   
\\
&&&&&&&&&&& $b_{12\, 12}$  
\end{tabular} \right), 
\\
D &= \left(
\begin{tabular} {ccrc crcc rccr}
0 & 0 & $d_{13}$ & 0 & 0 & $d_{16}$ & 0 & 0 & $d_{19}$ & 0 & 0 & $d_{1\, 12}$ 
\\ 
0 & 0 & $-d_{13}$ & 0 & 0 & $-d_{19}$ & 0 & 0 & $-d_{16}$ & 0 & 0 & $-d_{1\, 12}$ 
\\
0 & 0 & $d_{33}$ & 0 & 0 & $d_{36}$ & 0 & 0 & $-d_{36}$ & 0 & 0 & $d_{3\, 12}$ 
\\ 
0 & 0 & $d_{43}$ & 0 & 0 & $d_{46}$ & 0 & 0 & $d_{49}$ & 0 & 0 & $d_{4\, 12}$ 
\\
0 & 0 & $-d_{43}$ & 0 & 0 & $-d_{49}$ & 0 & 0 & $-d_{46}$ & 0 & 0 & $-d_{4\, 12}$ 
\\ 
0 & 0 & $d_{63}$ & 0 & 0 & $d_{66}$ & 0 & 0 & $-d_{66}$ & 0 & 0 & $d_{6\, 12}$ 
\\
0 & 0 & $d_{73}$ & 0 & 0 & $d_{76}$ & 0 & 0 & $d_{79}$ & 0 & 0 & $d_{7\, 12}$ 
\\ 
0 & 0 & $-d_{73}$ & 0 & 0 & $-d_{79}$ & 0 & 0 & $-d_{76}$ & 0 & 0 & $-d_{7\, 12}$ 
\\
0 & 0 & $d_{93}$ & 0 & 0 & $d_{96}$ & 0 & 0 & $-d_{96}$ & 0 & 0 & $d_{9\, 12}$ 
\end{tabular} \right). 
\end{align}

\subsubsection{SO(2) conditions on the elements of F}

We are now ready to apply  the SO(2) invariance condition \rf{113} to the simplified form of 
$F$ with 75 independent elements, where the matrix $\hat{P}$ defined by \rf{42} with ${\bd p}= {\bd e}_3$ has 36 nonzero elements.  The system of equations generated is large but relatively straightforward to solve, particularly with the aid of an electronic computer.  We simply state the results, which the reader may check. 

We find that the  matrix in \rf{113} vanishes if and only if  40 additional relations among the 75 independent elements are met.  These may be split into two sets, of 15 and 25 respectively.  The first set of 15  relations \rf{4021} involve only those elements of $F$ in which the index $1$ (and hence $2$) occurs an even number of times, and therefore they act only on the elements of $A_{9\times 9}$ and the non-zero elements of \rev{$B^{(1)}_{12\times 12}$} of \rf{4151}. 
The second set of 25 relations involve only elements of $F$ with the index $1$ (and  $2$) occurring an odd number of times:   
\begin{eqnarray}\label{423}
\begin{array}{rcrcrcr}
b_{1,2} = 0,&&   b_{4,8}=0  ,&& b_{5,7}= 0,&& b_{10,11}= 0,
 \\ %&&&&&& \\
 b_{1, 5}-b_{1, 11}  = 0 & ~~ &  b_{1, 5}  +b_{5, 10} = 0, &~~&  b_{1, 8}+b_{4, 5} -b_{4, 11}  =0, &~~&
\\ &&&&&& \\
d_{1,6} = 0,&&  d_{1,9}= 0 ,&& d_{3,3}= 0 ,&&  d_{3,6}= 0 , 
\\ %&&&&&& \\
d_{3,12}= 0,&&  d_{4,3}= 0 ,&& d_{6,3}= 0 ,&&  d_{6,6}= 0 ,  
 \\ %&&&&&& \\
d_{6,12}= 0,&& d_{7,12}= 0 ,&& d_{9,3}= 0 ,&&  d_{9,6}= 0 , 
 \\ %&&&&&& \\
d_{9,12}= 0,&&  d_{1,3}-d_{4,6}= 0 ,&&  d_{1,3}-d_{4,9}=0 ,&&  d_{1,12}-d_{7,6}=0 ,
\\ %&&&&&& \\
d_{1,12}-d_{7,9}= 0 ,&& d_{4,12} +d_{7,3}=0. &&   &&
\end{array}
\end{eqnarray}

Thus, 
\bal{456}
B = \left(
\begin{tabular} {cccc cccc cccc}
       $b_{11}$ & 0 & 0 & $b_{14}$ & $\bf b_{15}$& 0 &$b_{17}$     & $\bf b_{18}$& 0 & $b_{1\, 10}$     & $\bf b_{15}$ &  0
\\
&      $b_{11}$ & 0 & $-\bf b_{18}$ & $b_{17}$& 0 &$-\bf b_{15}$     & $b_{14}$& 0 & $\bf -b_{1\, 5}$     & $b_{1\, 10}$ &  0 
\\
& &    $b_{33}$ & 0& 0 & $b_{36}$ & 0& 0 &$b_{36}$     & 0& 0 & $b_{3\, 12}$  
\\
&&&    $b_{44}$ & $\bf b_{45}$& 0 &$b_{47}$     & 0 & 0 & $b_{4\, 10}$     & $\bf b_{1, 8}+b_{4, 5}$ &  0
\\
&&&&   $b_{55}$& 0 & 0     & $b_{47}$& 0 & $-\bf b_{15}$     & $b_{4\, 10}$ &  0 
\\
&&&&&  $b_{66}$ & 0& 0 & $b_{69}$ & 0& 0 &$b_{6\, 12}$  
\\
&&&&&&   $b_{55}$     & $-\bf b_{45}$& 0 & $b_{5\, 11}$     & $\bf b_{15}$ &  0
\\
&&&&&&&  $b_{44}$& 0 & $-\bf b_{1, 8}-b_{4, 5}$     & $b_{4\, 10}$ &  0
\\
&&&&&&&& $b_{66}$ & 0& 0 & $b_{6\, 12}$  
\\
&S&Y&M&&&&&& $b_{10\, 10}$     & 0 &  0 
\\
&&&&&&&&&&$b_{10\, 10}$ & 0
\\
&&&&&&&&&&& $b_{12\, 12}$  
\end{tabular} \right)
\end{align}
and $D$ is given by \rf{445}. 
The terms in bold in \rf{456} are essentially decoupled from the others, and we therefore split $F$ according to eq. \rf{4414} in order to emphasize the disjoint nature of the subspaces generated by the elements with index $1$ occurring an even and an odd number of times.  

The partition \rf{4414} also allows us to easily determine the dimensionality of $F$.  Thus, $F^{(2)}$ clearly has 6 independent elements. There are 44 distinct elements in $F^{(1)}$, and all but  4 of these, $f_{3\, 3}$, $f_{3\, 4}$, $f_{3\, 7}$ and 
$b_{4\, 4}$,  occur in the  relations \rf{4021}.  These 15 conditions are of rank 15, indicating that the conditions are linearly independent.  This can be verified by noting that each of the 15 equations involves at least one  element (the final one in the left member) not contained in the other 14 equations.  Hence, there are a total of 15 constraints on the 50 elements of $F$, which therefore has   50-15 = 35 independent elements.  This is the dimension of $F$, and the number of independent quadratic forms on $\tens{E}$la.

\subsection{SO(3) conditions on the elements of F}

There are various ways to deduce the SO(3) form of $F$ using the results for SO(2) (actually O(2)).  As discussed before, 
the required group of transformations is simply $O(3)$.  We also note the following as a consequence of Theorem \ref{thm2} and Lemma \ref{lem2}: 

\begin{lem}\label{lem3}
All quadratic invariants of $c$ under O(2)$\cup R({\bd e}_\perp)$ where ${\bd e}_\perp$ is perpendicular to the axis of rotation are of the form $c^tF^{(1)} c$ where the 21$\times$21 symmetric matrix $F^{(1)}$ is given by eq. $\rf{407}_1$.  

The number of linearly \rev{independent} quadratic invariants under SO(2)$\cup R({\bd e}_\perp)$ is 29.
\end{lem}

We can therefore start with $F^{(1)}$ of Theorem \ref{thm2}, which corresponds to  O(2)$\cup R({\bd e}_1)$.   The next step is to consider quarter turns about ${\bd e}_1$ and ${\bd e}_2$, which reduces $A$ and $B$ to the forms \rf{501} and \rf{5011}, respectively. 
The 12 distinct elements in $A$ and 7 in $B$ are related by the 15 conditions of Theorem \ref{thm2} in eqs. \rf{4021}.  We find that only 12 of these are linearly independent, or in other words, the system of equations is rank(12).   A linearly independent system can be obtained by, for instance, ignoring the final three conditions in \rf{4021}, to give a system of 12 equations on the 19 elements, i.e. eqs. \rf{505}.  

 In summary,  there are 12 relations between the 19 elements of   $F$.  Hence there are 7 quadratic invariants under SO(3). 
 
It remains to discuss these results with respect to the  invariants proposed by 
\citet{Ting87} and \cite{Ahmad02}.

\section{Consistency and completeness }  \label{sec6}

A given quadratic form can be checked to see if it is consistent with one of the invariant forms defined in Theorems \ref{thm1} and \ref{thm2}.  
We will describe how to do this and discuss the consistency of the quadratic invariants  proposed by Ting and Ahmad.  We note in passing that $A_1, A_2$ and $B_1, \ldots , B_4$ of eqs. \rf{066} and \rf{067} are obviously consistent with SO(3) invariance, but some of the SO(2) invariants proposed by \rev{Ting} and Ahmad are not immediately obvious.    We also discuss the completeness of Ting and Ahmad's quadratic invariants. 

\subsection{SO(3)}

\subsubsection{Consistency}
Given a quadratic form $\phi$ in $c$,  for instance $A_1^2$ of eq. \rf{066}, define a 21$\times$21 symmetric matrix  $F$ according to 
\beq{060}
f_{ij} = \frac12 \frac{ \partial^2 \phi} {\partial c_i\partial c_j}, 
\eeq
where $c_i$ is the $i-$th component of the modulus vector $c$ of \rf{00}.  The quadratic form $\phi$
is an invariant under SO(3) iff (i) the elements of $F$  have the form as defined by eqs. \rf{3414}-\rf{5011}, and (ii) they satisfy the 12 equations \rf{505}. 

It may be confirmed that Ahmad's  seven quadratic forms $\{A_1^2,\, A_2^2,\, A_1A_2,\,
B_1,\, B_2,\, B_3,\, B_4\}$ are  SO(3) invariants. 

\subsubsection{Completeness}

Assuming a quadratic form $\phi$ is consistent with SO(3) invariance, define  the vector of the 19 distinct elements in $F$ of \rf{501} and \rf{5011}, specifically   
\beq{1234}     % 1         2          3         4         5         6         7         8         9       10        
u=u_{19\times 1} = \big(  f_{11}\  f_{12}\  f_{14}\  f_{15}\  f_{17}\  f_{18}\  f_{44}\  f_{45}\  f_{47}\ f_{48}\ 
%
%\nonumber \\ %   11      12        13        14          15             16        17        18          19
             f_{77}\ f_{78}\   b_{11}\  b_{14}\  b_{1\, 10}\    b_{44}  \ b_{47}  \ b_{4\, 10} \ b_{10\, 10}
\big)^t . 
\eeq%nd{align}
Now suppose we have a set of different quadratic forms, $\phi_1,\ldots , \phi_N$, where $N\le 7$ (we do not need to consider $N>7$ since there can be no more than 7 linearly independent forms). Let $u_i$ be the 19-vector for $\phi_i$, $i=1,\ldots, N$, and define the matrix 
\beq{500}
M = M_{19\times N} = \big( u_1 \ u_2 \ \ldots \, u_N\big). 
\eeq
Then the quadratic forms are linearly independent iff rank$(M)=N$. By definition, a set of 7 quadratic forms is complete if they are linearly independent. 

It may be checked that Ahmad's  seven quadratic forms define a matrix $M$ of rank 7, and are therefore complete.

\subsection{SO(2)}

Checking a given quadratic form for consistency is analogous to the procedure described for SO(3).  Thus, first compute $F$ according to \rf{060}, then check 
that  (i) the elements of $F$  have the form as defined by eqs. \rf{3414}-\rf{5011}, and (ii) they satisfy the 15 equations \rf{4021}. 

It may be confirmed that Ahmad's  17  quadratic forms $E_1, \ldots , E_{17} $ in eq. \rf{9140} and  the new quadratic  forms  $E_{18}, E_{19} , E_{20} $ are all SO(2) invariants, as are the 15 quadratic  forms defined by the five linear invariants $L_1, \ldots , L_5 $ of eq. \rf{9103}. 

These 35 quadratic forms are also complete.  In order to see this define   
 $v=v_{50\times 1}$ as the vector of the 50 distinct elements in $F$ of \rf{415} - \rf{445}, specifically   
\bal{0234}     % 1         2          3         4         5         6         7         8         9        10        11
v = \big( & f_{11}\  f_{12}\  f_{13}\  f_{14}\  f_{15}\  f_{16}\  f_{17}\  f_{18}\  f_{19}\  f_{33}\  f_{34}\   f_{36}\  f_{37}\ f_{39}\  f_{44}\ f_{45}\ f_{46}
\nonumber \\ %   12        13       14        15       16       17       18       19       20       21        22
  &           f_{47}\ f_{48}\ f_{49}\ f_{66}\  f_{67}\   f_{69}\  f_{77}\ f_{78}\ 
    f_{79}\ f_{99}\  b_{11}\ b_{14}\ b_{15} \  b_{17}\ b_{18}\  b_{1\, 10}\  b_{33}\
\nonumber \\ %   23       24       25       26       27       28        29       30         31         32        33
&             b_{36}\ 
           b_{3\,12}\ b_{44}\  b_{45}\  b_{47}\ b_{4\,10}\ b_{55}\ b_{5\,11}\ b_{66}\ b_{69}\ b_{6\,12}\ b_{10\,10}\ b_{12\,12}\ d_{13}\  d_{1\, 12}\  d_{4\, 12} \big)^t . 
\end{align}
Let $M$ be the matrix formed from the  $v$-vectors of the  35 quadratic forms.  Then it may be checked that rank$(M)=35$, indicating that they are a complete set. 

Finally, we note that 
%\beq{563}   w^t = \big(  b_{15}\  b_{18}\  b_{45}\   d_{13}\  d_{1\, 12}\  d_{4\, 12} \big) . 
%\eeq
\beq{790}
\frac12 c^tF^{(2)}c = -d_{4\, 12} E_{15} -d_{13} E_{16} +d_{1\, 12} E_{17} 
+b_{15} E_{18} +(b_{18}+b_{45}) E_{19} - b_{18} E_{20}.
\eeq
This illustrates how the partition of $F$ in eq. \rf{4414} splits the set of invariants into distinct subsets, one associated with elements of $F$ that have index $1$ (and $2$) an even number of times, which is a 29 dimensional subspace.  The other is the six dimensional subspace associated with the fact that SO(2) is not a subspace of  orthorhombic symmetry for this type of 8$^{th}$ order tensor.

%%%%%%%%%%%%%%%%%%%%%%%%%%%%%%%%%%%%%%%%%%%%%%%%%%%%%%%%%%%%%%%%%%%%%%%%%%
\appendix  % only once

\Appendix{Proof of equation \rf{081}}
Define the totally symmetric and the asymmetric parts of the elasticity tensor as 
\beq{-101}
C^{(s)}_{ijkl} = \frac13 ( C_{ijkl}+C_{ikjl}+C_{iljk} ),  
\qquad
C^{(a)}_{ijkl} = \frac13 ( 2C_{ijkl}-C_{ikjl}-C_{iljk} ),
\eeq
respectively.  These partition the elastic moduli,  
\beq{-12}
C_{ijkl} = 
C^{(s)}_{ijkl} +
C^{(a)}_{ijkl} ,
\eeq
and are orthogonal in the sense that 
\beq{-13}
C_{ijkl}C_{ijkl} = 
C^{(s)}_{ijkl}C^{(s)}_{ijkl} +
C^{(a)}_{ijkl}C^{(a)}_{ijkl}. 
\eeq
Hence, 
\bal{-14}
2C_{ijkl}C_{ikjl} &=  C_{ijkl}(C_{ikjl} + C_{iljk} )
\nonumber \\
&=  C_{ijkl} ( 3C^{(s)}_{ijkl} - C_{ijkl} )
\nonumber \\
&=  C_{ijkl} ( 2C^{(s)}_{ijkl} - C^{(a)}_{ijkl} )
\nonumber \\
&=  2 C^{(s)}_{ijkl}C^{(s)}_{ijkl} - C^{(a)}_{ijkl}C^{(a)}_{ijkl}
\nonumber  \\
&=  2 C_{ijkl}C_{ijkl} - 3 C^{(a)}_{ijkl}C^{(a)}_{ijkl}, 
\end{align}
and therefore
\beq{-15}
B_5 = B_1 - \frac32 C^{(a)}_{ijkl}C^{(a)}_{ijkl}.
\eeq

While it is easy to verify that 
\beq{-16}
 C^{(a)}_{ijkk} = \frac23 D_{ij}, 
\eeq
where 
\beq{-17}
D_{ij}  =  C_{ijkk} - C_{ikjk},   
\eeq
it is not as obvious that $C^{(a)}$ is completely defined by $D$.  It may be shown  \citep{backus,Norris05g} that 
\bal{ca}
C^{(a)}_{ijkl}=&\frac13\big[ 2D_{ij}\delta_{kl}+ 2D_{kl}\delta_{ij} 
- D_{ik}\delta_{jl}- D_{il}\delta_{jk} - D_{jk}\delta_{il} - D_{jl}\delta_{ik}
\nonumber \\
 & \quad 
+ \frac12  D_{mm} (\delta_{ik}\delta_{jl}+ \delta_{il}\delta_{jk} - 2 \delta_{ij}\delta_{kl} )
\big], 
\end{align}
and hence, 
\beq{-29}
C^{(a)}_{ijkl}C^{(a)}_{ijkl} = \frac43 D_{ij}D_{ij} - \frac13 D_{ii}D_{jj}. 
\eeq
Finally, 
\beq{845}
B_5 = B_1 + \frac12 D_{ii}D_{jj} - 2 D_{ij}D_{ij} , 
\eeq
from which eq. \rf{081} follows.

%%%%%%%%%%%%%%%%%%%%%%%%%%%%%%%%%%%%%%%%%%%%%%%%%%%%%%%%%%%%%%%%%%%%%%%%%%
%\bibliography{../bib/thermoelastic}

\end{document}